\documentclass[aip,reprint]{revtex4-1}

\usepackage{graphicx}
\usepackage{amsmath}
\usepackage{amssymb}
\usepackage{dcolumn}
\usepackage{bm}
\usepackage{hyperref}
\usepackage{makecell}

\begin{document}


\title{X-ray Spectroscopy of Hot Dense Plasmas Using LLNL's Upgraded
  Titan 2$\omega$ Short-Pulse Laser}

\author{A.~J.~Fairchild}
\email{fairchild8@llnl.gov}
\affiliation{Lawrence Livermore National Laboratory, Livermore, CA 94550, USA}

\author{D.~T.~Bishel}
\affiliation{Lawrence Livermore National Laboratory, Livermore, CA 94550, USA}

\author{T.~Cordova}
\affiliation{University of California San Diego, La Jolla, CA 92093, USA}

\author{H.~Flores-Alimboyoguen}
\affiliation{Lawrence Livermore National Laboratory, Livermore, CA 94550, USA}

\author{M.~F.~Gu}
\affiliation{Prism Computational Sciences, Madison, WI, 53711, USA}

\author{P.~Hamilton}
\affiliation{Lawrence Livermore National Laboratory, Livermore, CA 94550, USA}

\author{N.~Hell}
\affiliation{Lawrence Livermore National Laboratory, Livermore, CA 94550, USA}

\author{M.~J.~MacDonald}
\affiliation{Lawrence Livermore National Laboratory, Livermore, CA 94550, USA}

\author{E.~W.~Magee}
\affiliation{Lawrence Livermore National Laboratory, Livermore, CA 94550, USA}

\author{R.~C.~Mancini}
\affiliation{University of Nevada, Reno, Reno, NV 89557, USA}

\author{E.~V.~Marley}
\affiliation{Lawrence Livermore National Laboratory, Livermore, CA 94550, USA}
\affiliation{Laboratory for Laser Energetics, Rochester, NY 14623, USA}

\author{R.~Shepherd}
\affiliation{Lawrence Livermore National Laboratory, Livermore, CA 94550, USA}

\author{G.~V.~Brown}
\affiliation{Lawrence Livermore National Laboratory, Livermore, CA 94550, USA}

\begin{abstract}
We are developing an experimental platform at Lawrence Livermore National Laboratory's Jupiter Laser Facility to characterize X-ray emission from highly charged ions in the high-density ($n_e > 10^{23}$~cm$^{-3}$), high-temperature ($T_e \sim 1$~keV) regime, leveraging the recently upgraded frequency-doubled short-pulse Titan laser. These measurements will provide experimental benchmarks for high electron density effects on atomic structure, including Stark broadening, ionization potential depression, density-dependent line shifts, and suppression of dielectronic recombination, effects that are relevant to spectral modeling packages used by both the high-energy-density and astrophysics communities. Here, we describe the current status of the platform and present results from first experiments using the upgraded frequency-doubling capability, including calibrated K-shell spectra acquired from five time-integrated X-ray channels. Comparisons with MERL line shape calculations and with spectra recorded at AWE's Orion Laser Facility indicate that the upgraded Titan laser reliably produces the necessary plasma conditions, establishing a foundation for systematic benchmarking of high-density atomic physics in future campaigns.
\end{abstract}

\maketitle

\section{\label{sec:intro}Introduction}

In high-density plasmas, the electric fields generated by neighboring ions and free electrons modify atomic energy levels, leading to effects such as ionization potential depression~\cite{Ciricosta2012, Hoarty:PRL:2013}, density-dependent line shifts~\cite{beiersdorfer:PRA:2019}, collisional quenching of metastable states~\cite{Porquet:AASS:2000}, and dielectronic recombination suppression~\cite{Nikolic:APJ:2013}. These changes alter the observed spectral lines and line shapes and provide valuable diagnostics for extreme astrophysical and laboratory plasmas. Accurately including these effects into plasma models is required to reliably fit and interpret the measured spectra. However, progress is limited by the lack of experimental benchmarks of models (and model components) in the high-density regime. 

High-density atomic physics is particularly important for interpreting X-ray reflection spectroscopy of black hole accretion disks, where inflated iron abundances have been linked to the neglect of these effects in current models~\cite{garcia2018a}. In the extreme environments near black holes, accretion disk densities can reach $10^{18}$--$10^{23}$~cm$^{-3}$, regimes where high-density effects significantly alter atomic structure and K-shell X-ray emission. Current reflection spectroscopy analyses of both active galactic nuclei (AGN) and black hole X-ray binaries (BHBs) systematically infer iron abundances several times solar, with no clear astrophysical enrichment mechanism to explain these values~\cite{garcia2018a}. Moreover, a strong correlation between inferred iron abundance and black hole spin has been identified, suggesting that model systematics---likely due to neglected high-density effects---may bias spin estimates towards higher absolute spin values~\cite{garcia2015a}. Recent advances in reflection models have begun incorporating high-density effects, bringing inferred iron abundances closer to solar values \cite{tomsick2018,jiang2018, mallick2018, jiang2019, garcia2019b}; however, progress is limited by the absence of laboratory benchmarks at the relevant densities and temperatures.

These same high-density effects are central to high-energy-density (HED) science. Direct measurements of ionization potential depression~\cite{Ciricosta2012, Hoarty:PRL:2013} and its effect on emitted spectra~\cite{Preston2013} have revealed long-standing disagreement between experiment and widely used plasma-screening models. In addition, systematic code-comparison studies have shown that such density-dependent atomic physics can introduce significant uncertainty into the modeling of HED experiments~\cite{Hansen2020}. Resolving these discrepancies, including for the design and interpretation of inertial confinement fusion (ICF) experiments~\cite{Gaffney2013}, requires experimental benchmarks across a broader range of elements, densities, and temperatures than currently exist.

Laboratory laser experiments provide a route to generate these conditions in a controlled environment where plasma parameters can be independently diagnosed. The Jupiter Laser Facility's (JLF) Titan laser at Lawrence Livermore National Laboratory (LLNL) recently completed an extensive refurbishment, returning to full operation at the end of 2023. Among the upgrades is the addition of full-aperture frequency doubling on the short-pulse beam to $\lambda = 527$~nm (2$\omega$) achieved via a potassium dihydrogen phosphate (KDP) crystal, delivering up to 50~J on target in pulses ranging from 0.7--200~ps. The frequency doubling significantly improves the temporal pulse contrast, which is essential for these experiments: by suppressing the prepulse that would otherwise ablate and expand the tamping layers prematurely, the 2$\omega$ beam enables production of a more uniform, well-confined plasma at the required temperatures and densities.

In this paper, we present first results from the upgraded Titan 2$\omega$ short-pulse laser using a suite of time-integrated X-ray spectrometers. We demonstrate that the JLF/Titan 2$\omega$ short-pulse laser can produce plasma conditions comparable to those achieved at the Atomic Weapons Establishment (AWE) Orion Laser Facility, establishing a robust platform for systematic benchmarking of high-density atomic physics relevant to both the HED and astrophysics communities.

\section{\label{sec:methods}Experimental Setup}

The Titan laser at LLNL's JLF has a dedicated short-pulse beamline and a second long-pulse beamline, shared with the Janus laser. The frequency-doubled (2$\omega$) short-pulse beam delivers up to 50~J at 527~nm with pulse durations of 0.7--20~ps. 
For these experiments, the short-pulse beam was focused using a f/3 off-axis parabola to achieve irradiances of $10^{17}$--$10^{19}$~W~cm$^{-2}$. The frequency-doubled long-pulse  beam operates at 527~nm over 0.5--20~ns pulse durations with energies up to 500~J. Long-pulse shots were used exclusively during the initial spectrometer setup. Due to the delayed arrival of the newly coated focusing parabola designed for 2$\omega$, only seven short-pulse shots (labeled shots 46--52) were obtained with laser energies ranging from $\sim$ 20--50~J and with focal spot sizes estimated between 15 and $>100~\mu$m. For these commissioning experiments of Titan's frequency-doubled capability, shot-to-shot measurements of the laser spot size were not available. 

KCl buried-layer foil targets were the primary short-pulse targets, with a target stack $\sim$ 2~mm $\times$ 2~mm in extent. The KCl layers are buried between parylene-N tampers to confine the ablation plasma and maintain high densities, with thin carbon layers providing adhesion of the KCl to the parylene-N. From the laser irradiation side, the stack consisted of nominally 3.0~$\mu$m parylene-N, 0.03~$\mu$m Carbon, 0.15~$\mu$m KCl, 0.03~$\mu$m Carbon, and 3.0~$\mu$m parylene-N. Iron foil targets consisting of 74~$\mu$m Fe foil coated on both sides with 1~$\mu$m of parylene-D and slurry targets \cite{Smith:JAP:2022} of NaCl and KCl/CaHPO$_4$ were used during the long-pulse setup phase to help setup the spectrometers. KCl buried-layer targets with an additional layer of 0.15~$\mu$m Fe were purchased as well, but were not used due to delamination prior to building the targets.

\begin{figure*}[htbp]
  \centering
  \includegraphics[width=\textwidth]{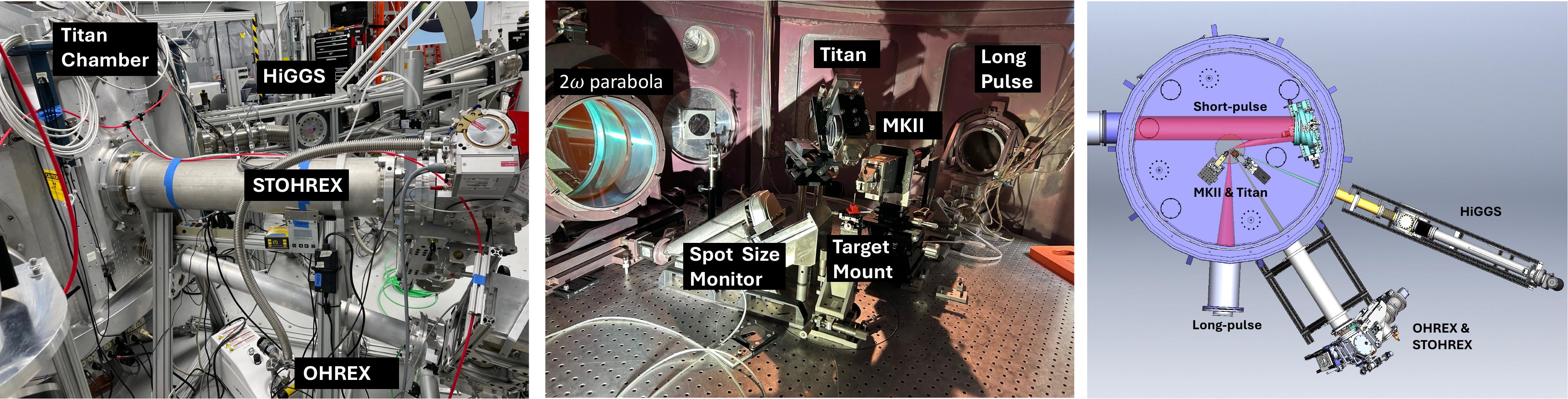}
  \caption{Left: photo of the outside of the Titan chamber showing the attached HiGGS, OHREX, and STOHREX spectrometers. Middle: photo inside the Titan chamber displaying the MKII and Titan spectrometers, the target mount, and the 2$\omega$ focusing parabola. Right: eDrawing of the experimental setup.}
  \label{fig:exp}
\end{figure*}

Five time-integrated X-ray spectrometers were fielded, covering seven spectral channels for temperature, density, and charge-balance diagnostics. Table~\ref{tab:diagnostics} summarizes the suite, and Figure~\ref{fig:exp} shows the experimental setup, including photographs of the spectrometers mounted on the Titan target chamber and an schematic of their relative positions and viewing geometry. For the short-pulse shots, the diagnostics had the following viewing angles. The polar angle $\theta$ is measured from the target normal in the target-chamber midplane, with positive angles measured clockwise and negative angles measured counterclockwise. The azimuthal angle $\phi$ specifies the out-of-plane angle measured from the target-chamber midplane about target chamber center (TCC); unless otherwise stated, the spectrometer was located in the midplane, such that $\phi=0$. The short-pulse lines of sight were HiGGS, $\theta=-20^\circ$; STOHREX, $\theta=+15^\circ$; OHREX, $\theta=+15^\circ$ and $\phi=-15^\circ$; MKII, $\theta=+35^\circ$; and Titan, $\theta=+11^\circ$, $\phi = 35^\circ$.

The MKII, Titan, and OHREX spectra were recorded on Fuji BAS-TR image plates, while HiGGS used a back-illuminated CCD detector. The image plates were scanned on a Fuji FLA7000 flatbed scanner with a $25~\mu\mathrm{m}$ scan pitch and saved as logarithmic grayscale images. Based on the measured BAS-TR spatial response reported by Fiksel \textit{et al.}~\cite{Fiksel:RSI:2012}, we adopt a rounded $100~\mu\mathrm{m}$ effective detector-plus-scanner spatial-response element, corresponding to approximately four raw scanner pixels. For each image-plate spectrometer, the local calibrated dispersion was used to convert this spatial scale into an effective detector-response width, $\Delta E_{\rm det}$, and a detector-response-limited resolving power, $R_{\rm det}=E/\Delta E_{\rm det}$. These values represent approximate upper scales for the resolving power of the experimental configuration. Resolving powers obtained by fitting measured spectral features, $R_{\rm obs}$, include these instrumental contributions together with Doppler broadening, Stark broadening, and potentially unresolved line structure and therefore provide lower bounds on the instrumental resolving power. No correction for image-plate fade between exposure and scanning was applied, and no background correction has been applied to the spectra presented here.

H- and He-like reference energies for determining spectrometer energy scales were taken from the curated reference tables of Hell \textit{et al.}~\cite{Hell:arXiv:2025}, whose tabulated values are taken from the calculations of Yerokhin and Surzhykov~\cite{Yerokhin:JPCRD:2015, Yerokhin:JPCRD:2019}. The Ne-like Fe line energies used to calibrate the HiGGS were taken from Brown \textit{et al.}~\cite{Brown_1998}. Where appropriate we utilize the letter designations for K$\alpha$ transitions in He- and Li-like ions of Gabriel \cite{Gabriel:1972} and for L-shell transitions in Ne-like ions of Parkinson \cite{Parkinson:AAP:1973}.

We report calibrated spectra in units of time-integrated spectral intensity (J/keV/sr) by correcting for the X-ray response of the image plate, filter transmission, and the throughput of each spectrometer. We convert the logarithmic grayscale images to physical PSL/pixel (photo-stimulated luminescence) units using the formula from Maddox \textit{et al.}~\cite{Maddox:RSI:2011} and then to photons/pixel using the TR IP sensitivity calibration of Meadowcroft \textit{et al.}~\cite{Meadowcroft:RSI:2008}. Filter transmissions were computed using the mass densities and mass attenuation coefficients from \textit{xraylib}~\cite{Kissel:RPC:2000, Elam:RPC:2002, Schoonjans2011}. The throughput of each spectrometer was determined from the integrated crystal reflectivity together with the spectrometer collection efficiency. The collection efficiency includes both the solid angle subtended by the crystal and the geometric redistribution of that solid angle onto the detector by the crystal's focusing properties. These geometric factors were determined using ray tracing simulations. For the The Orion high-resolution X-ray spectrometer (OHREX) spectrometer, the ray tracing analysis was done using \textit{XICSRT}~\cite{Pablant:RSI:2021, Cordova:RSI:2022} which showed similar results to MacDonald \textit{et al.}~\cite{MacDonald:RSI:2021}. For the MKII/Titan spectrometers, a custom ray tracing analysis was built in Python and was checked against the \textit{XICSRT} results. The integrated reflectivities of each crystal were calculated using \textit{pyTTE}~\cite{Honkanen:hf5943}. The absolute scale of the intensity calibration will be improved in future work through measurements of the integrated reflectivities, detector responses, and filter thicknesses~\cite{Park:RSI:2010, MacDonald:RSI:2021}.

\begin{table*}[htbp]
\caption{\label{tab:diagnostics}Summary of X-ray spectrometer diagnostics.}
\begin{ruledtabular}
\begin{tabular}{lcccc}
Diagnostic & Dispersive element & Energy range (keV) & Observes & Primary goal \\
\hline
MKII  & CsAP & 2.1--5.0  & Cl \& K K-shell & Time-integrated $T_e$ \& $n_e$ \\
Titan & \makecell[tc]{Quartz(100)\\Ge(111)} & \makecell[tc]{2.8--4.8\\2.5--3.2} & \makecell[tc]{Cl He$\beta$ \& K K-shell\\Cl He$\alpha$ \& Ly$\alpha$} & Time-integrated secondary diagnostic \\
OHREX & \makecell[tc]{Quartz(110)\\Quartz(110)} & \makecell[tc]{3.2--3.4\\3.3--3.6} & \makecell[tc]{Cl He$\beta$ \& Fe He$\alpha$ (2$^{nd}$ order)\\K He$\alpha$ \& Fe He$\alpha$ (2$^{nd}$ order)} & Time-integrated\ $T_e$ \& $n_e$ \\
HiGGS     & 2400~l/mm grating         & 0.77--1.24   & Fe L-shell                                & Time-integrated charge balance \\
STOHREX   & Quartz(110)               & 3.4--3.6  & K He$\alpha$; Fe He$\alpha$ (2$^{nd}$ order)   & Time-resolved\ $T_e$ \& $n_e$ \\
\end{tabular}
\end{ruledtabular}
\end{table*}

The MKII is a meridionally curved, convex cylindrical crystal spectrometer with an adjustable Bragg angle that provides broadband, time-integrated K-shell X-ray spectra. These K-shell spectra enable temperature estimates from the intensity ratios of He$\alpha$ to Ly$\alpha$ complexes and density estimates from the width of the optically thin, Stark-broadened He$\beta$ lines. For these experiments, we used a 500~mm radius of curvature CsAP crystal located 125~mm from TCC, with an optical blocking/debris filter stack comprised of 0.78~$\mu$m of Al, 19.5~$\mu$m of Mylar, 4.0~$\mu$m of polypropylene, 2.5~$\mu$m  of polyphenylene ether, and 0.2~$\mu$m of polycarbonate. Three of these filters were positioned upstream of the crystal at 44, 62.5, and 98.5~mm from TCC, where they also served as debris shields; the remaining two were positioned downstream of the crystal at 228 and 232.5~mm from TCC, the latter mounted on the image-plate holder.

Near the K He$\alpha$ feature at $3508~\mathrm{eV}$, the calibrated MKII dispersion was $\sim 2.38~\mathrm{eV}$ per $25~\mu\mathrm{m}$ raw scanner pixel. The adopted $100~\mu\mathrm{m}$ detector response therefore corresponds to $\Delta E_{\rm det}\simeq9.5~\mathrm{eV}$ and a detector-response-limited resolving power of $R_{\rm det}\simeq370$. A Gaussian fit to the measured K He$\alpha$ feature gave an observed resolving power of $R_{\rm obs}=257$. 

The Titan spectrometer is also a meridionally curved, convex cylindrical crystal spectrometer with an adjustable Bragg angle that houses two crystals and provides secondary, time-integrated temperature and density diagnostics. For these experiments, we used a Quartz(100) and a Ge(111) crystal, each with a 500~mm radius of curvature, located 145~mm from TCC, along with an optical blocking/debris filter stack comprised of 0.45~$\mu$m of Al and 19.5~$\mu$m of Mylar. All three filters were positioned upstream of the crystal at 83, 92, and 100~mm from TCC, where they also served as debris shields.

For the Ge(111) channel, the calibrated dispersion near the Cl He$\alpha$ feature at $2790~\mathrm{eV}$ was $\sim 0.48~\mathrm{eV}$ per raw scanner pixel. The adopted detector response corresponds to $\Delta E_{\rm det}\simeq1.9~\mathrm{eV}$ and $R_{\rm det}\simeq1470$. A Gaussian fit to the measured feature gave an observed resolving power of $R_{\rm obs}=490$.

For the quartz(100) channel, the calibrated dispersion near the K He$\alpha$ feature at $3510~\mathrm{eV}$ was $\sim 1.4~\mathrm{eV}$ per raw scanner pixel. The corresponding detector-response width is $\Delta E_{\rm det}\simeq5.38~\mathrm{eV}$, giving $R_{\rm det}\simeq650$. A Gaussian fit to the measured feature gave an observed resolving power of $R_{\rm obs}=514$.

The OHREX~\cite{beiersdorfer2016a,beiersdorfer2016b} is a spherically bent crystal spectrometer that houses two crystals and provides high-resolution K-shell spectra. The crystals have a radius of curvature of 672~mm and are a distance of 240 cm from the source. OHREX operates at a fixed central Bragg angle of 51.3\textdegree with a nominal cross-dispersion magnification of $M = 0.21$, and different wavelength ranges are chosen by selecting different crystals \cite{MacDonald:RSI:2021}. 

\begin{figure*}[ht]

  \centering
  \begin{minipage}{0.48\textwidth}
    \centering
    \includegraphics[width=\linewidth]{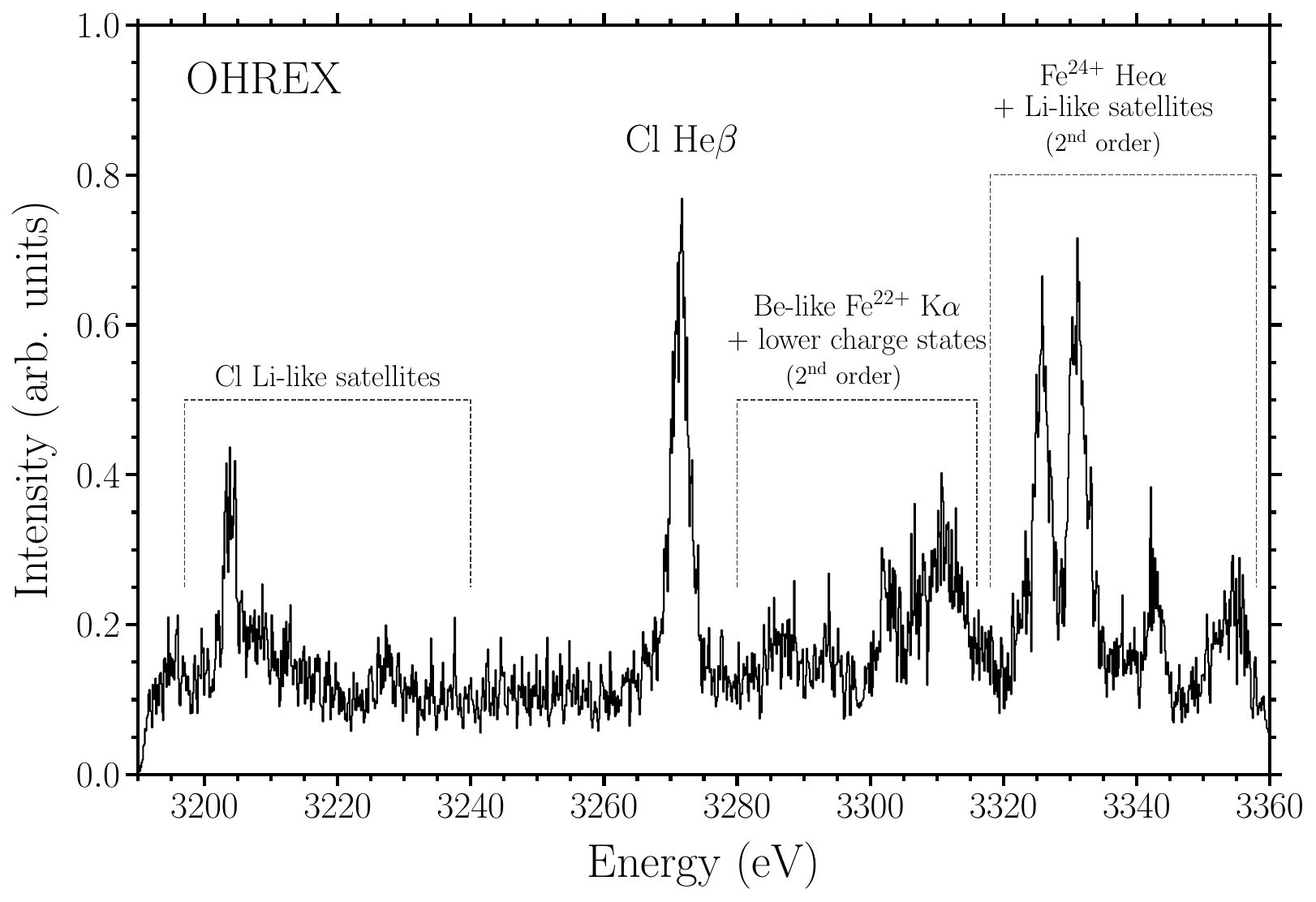}
  \end{minipage}
  \centering
  \begin{minipage}{0.48\textwidth}
    \centering
    \includegraphics[width=\linewidth]{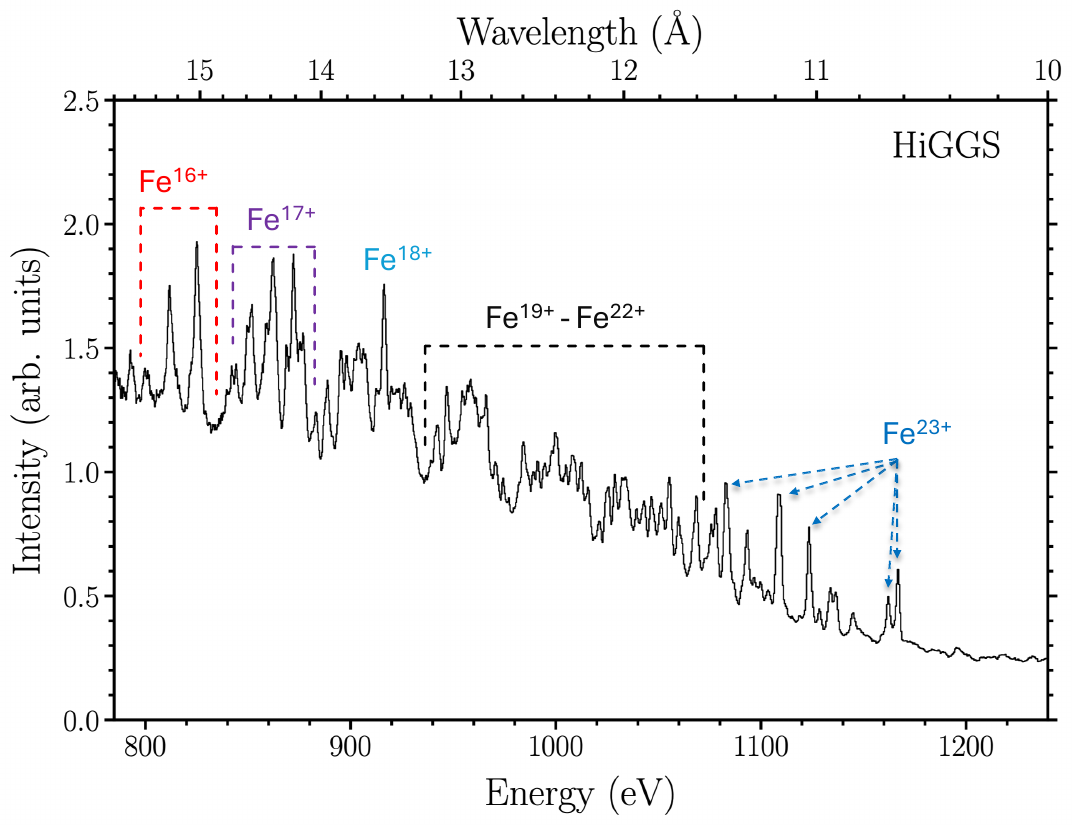}
  \end{minipage}\hfill
  \caption{Left: OHREX spectrum of Cl$^{15+}$ He$\beta$ and its Li-like satellites in first order, and Fe$^{24+}$ He$\alpha$ and its Li-like satellites in second order, produced by the long-pulse laser. Right: HiGGS Fe-L spectrum from the same shot. The ability to measure the K and L-shell spectra simultaneously constrains the iron charge-state distribution.}
  \label{fig:longpulse}
\end{figure*}

For these experiments, two Quartz(110) crystals were used with a 50.8~$\mu$m Be window. Quartz crystals have an intrinsic resolving powers, $ E / \Delta E > 10000$~\cite{Burek:SSI:1976}. For our experimental setup, the energy dispersion is $\sim 0.13~\mathrm{eV}$ per raw scanner pixel. The corresponding detector-response width is $\Delta E_{\rm det}\simeq0.5~\mathrm{eV}$, giving $R_{\rm det}\ge6500$. The narrowest fitted spectral feature in Figure~\ref{fig:longpulse} had an observed resolving power of $R_{\rm obs}=1600$. For comparison, OHREX measurements of the Cl He$\beta$ line at LLNL's electron beam ion trap (EBIT-I) facility using a Qz(110) crystal and CCD with 20~$\mu$m pixels, had measured resolving powers of approximately $R_{\rm obs}=3500$, where the measured resolving power was limited by the Doppler broadening of the $\sim$ 0.5 keV Cl ions~\cite{Hell:RSI:2016}.

The high-resolution grazing-incidence grating spectrometer \citep[HiGGS]{Beiersdorfer:RSI:2014} uses a 2400~l/mm reflection grating to record extreme ultraviolet (EUV) and soft X-ray spectra between 10--300~\AA, with the achievable resolving power depending on wavelength, source size, focus, and detector response~\cite{Beiersdorfer:RSI:2004,Beiersdorfer:RSI:2014}. The HiGGS typically operates with a liquid-nitrogen-cooled, back-illuminated CCD detector with 1340 $\times$ 1300 pixels of 20\,$\mu$m $\times$ 20\,$\mu$m area per pixel. For a given spectrometer setup, only $\sim$ 6~\AA{} near the short-wavelength end, or as much as $\sim$ 20~\AA{} near the long-wavelength end, are captured on the CCD. 

For these experiments, the HiGGS spectrometer was configured to observe the Fe L-shell spectrum from $\sim$ 10--16~\AA{} ($\sim$ 770--1240~eV) with an optical blocking/debris filter stack consisting of 8.0~$\mu$m of polypropylene and 0.04~$\mu$m of Al. Near the $n=3 \rightarrow 2$ Li-like Fe$^{23+}$ lines at $10.63~\mathrm{\AA}$, the calibrated wavelength scale was approximately $4.79~\mathrm{m\AA}$ per $20~\mu\mathrm{m}$ CCD pixel. Treating the approximately $25$--$36~\mu\mathrm{m}$ spatial-response scales reported for soft-x-ray CCDs as effective detector widths gives an estimated detector-response width of approximately $6.0$--$8.6~\mathrm{m\AA}$, corresponding to a detector-response-limited resolving power, $R_{\rm det}$, of approximately 1200--1800~\cite{Dinardo:NIMA:2007,Amorese:NIMA:2019}. A fit to narrowest of the Li-like Fe L-shell lines in Figure \ref{fig:longpulse} gave $R_{\rm obs}=590$, comparable to what similar spectrometers at the EBIT-I facility have achieved in this wavelength range~\cite{Beiersdorfer:RSI:2004}. The simultaneous measurement of time-integrated K-shell (OHREX, MKII, Titan) and L-shell (HiGGS) emission provides strong constraints on the overall charge balance. 

The streaked OHREX (STOHREX) \cite{Magee:RSI:2025} is the time-resolved counterpart to OHREX. Unfortunately, due to the delay in receiving the 2$\omega$ parabola, only time-integrated data on long-pulse setup shots with image plate were collected on the STOHREX.

OHREX and HiGGS were originally developed at LLNL's EBIT-I facility~\cite{Park:RSI:2010,Beiersdorfer:RSI:2014,Hell:RSI:2016,MacDonald:RSI:2021}. EBIT-I uses a magnetically compressed electron beam to produce and confine highly charged ions and can generate radiation from nearly any desired element and charge state, making it well suited for atomic-physics measurements and spectrometer development and calibration~\cite{Levine:PS:1988}. Prior to installation at JLF/Titan, both spectrometers were configured and tested at EBIT-I, confirming their intended spectral ranges and operation. This pre-campaign preparation substantially reduced technical risk during a shot-limited experiment. 

\begin{figure}[htbp]
  \centering
  \includegraphics[width=\columnwidth]{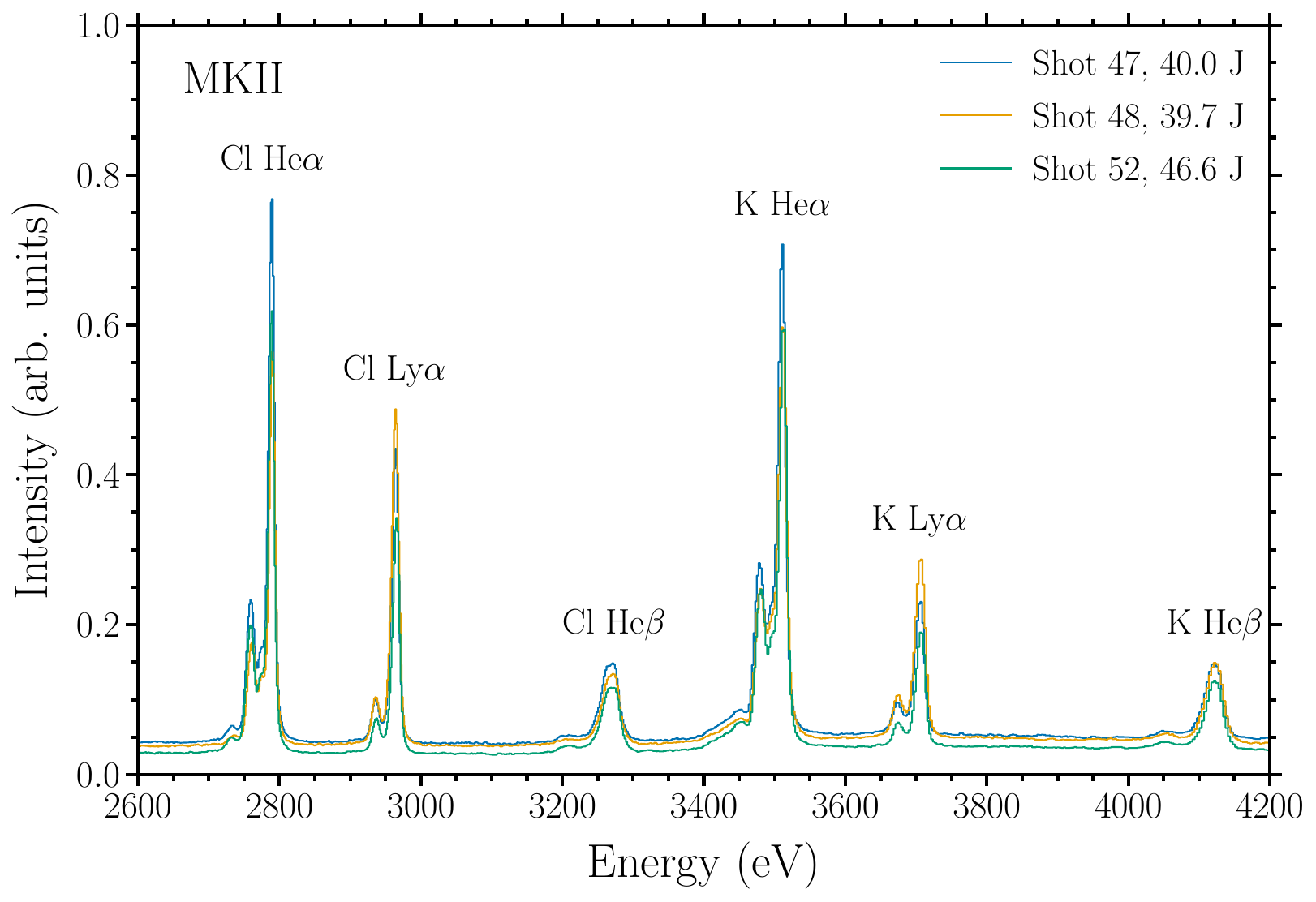}
  \caption{Comparison of raw MKII K-shell lineouts from three short-pulse shots (47, 48, and 52) on KCl buried-layer foil targets, with similar laser energies ($\sim$40~J) and pulse lengths (0.7~ps). The close agreement among shots demonstrates reproducible plasma conditions under similar laser parameters.}
  \label{fig:mkii_compare}
\end{figure}

\section{\label{sec:results}Results}

Figure~\ref{fig:longpulse} shows high-resolution OHREX and HiGGS spectra from long-pulse shots on a 74~$\mu$m Fe target coated with 1~$\mu$m parylene-D on both sides (shot 30). These are raw lineouts, with only corrections for the HiGGS filter transmission. The laser parameters were 214.5~J with a 12~ns pulse length. The OHREX spectrum shows the Cl$^{15+}$ He$\beta$ line and its Li-like satellites, arising from chlorine present in the parylene-D coating, in first order, and the Fe$^{24+}$ He$\alpha$ complex, its Li-like satellites, and K$\alpha$ transitions from lower charge states of Fe in second order. The ability to simultaneously measure the K-and L-shell transitions from the same plasma provides strong, self-consistent constraints on the electron temperature and charge balance. The OHREX energy scale was set by anchoring the XICSRT ray tracing determined dispersion to the Cl He$\beta$ line position. The HiGGS observes the Fe L-shell transitions, resolving charge states from Ne-like Fe$^{16+}$ through Li-like Fe$^{23+}$. The HiGGS wavelength scale was determined by fitting a second-order polynomial to the pixel-vs-wavelength values using Ne-like Fe$^{16+}$ 3G, 3F, 3D, 3C and He-like Na line w.

Figure~\ref{fig:mkii_compare} shows three MKII K-shell spectra from short-pulse shots on KCl buried-layer targets (shots 47, 48, and 52) with laser energies $\sim$ 40~J and 0.7~ps pulse lengths. These are raw lineouts, prior to corrections for image plate response, filter transmission, or spectrometer throughput. The energy scales were determined by fitting a third-order polynomial to pixel-versus-reference-energies using Cl, K He$\alpha$ w, Cl, K Ly$\alpha$, and Cl, K He$\beta$ lines. The close agreement among the three spectra indicates that similar laser parameters reproducibly generate similar plasma conditions.

\begin{figure*}[ht]

  \centering
  \begin{minipage}{0.48\textwidth}
    \centering
    \includegraphics[width=\linewidth]{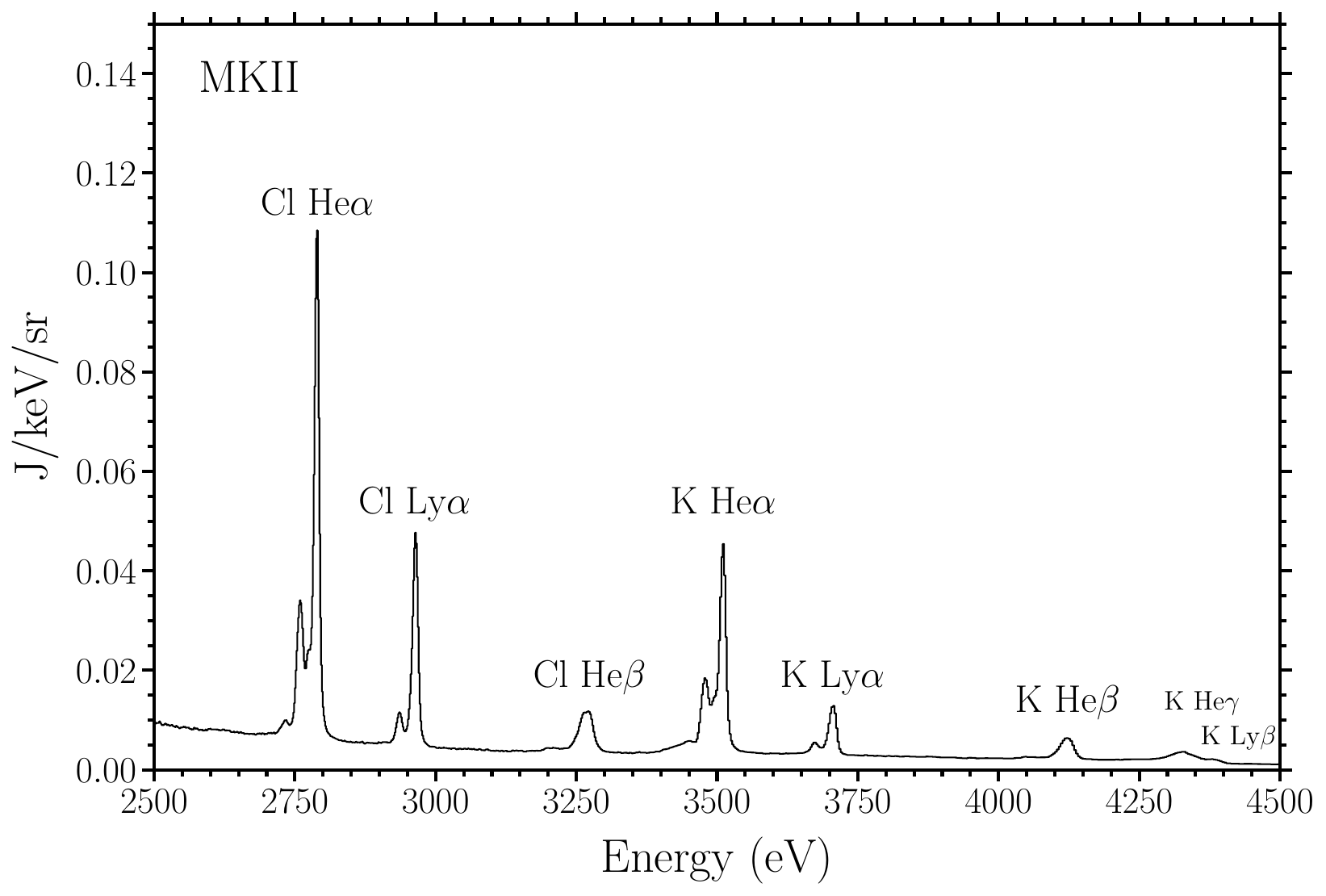}
  \end{minipage}
  \vspace{0.8em}

  \centering
  \begin{minipage}{0.48\textwidth}
    \centering
    \includegraphics[width=\linewidth]{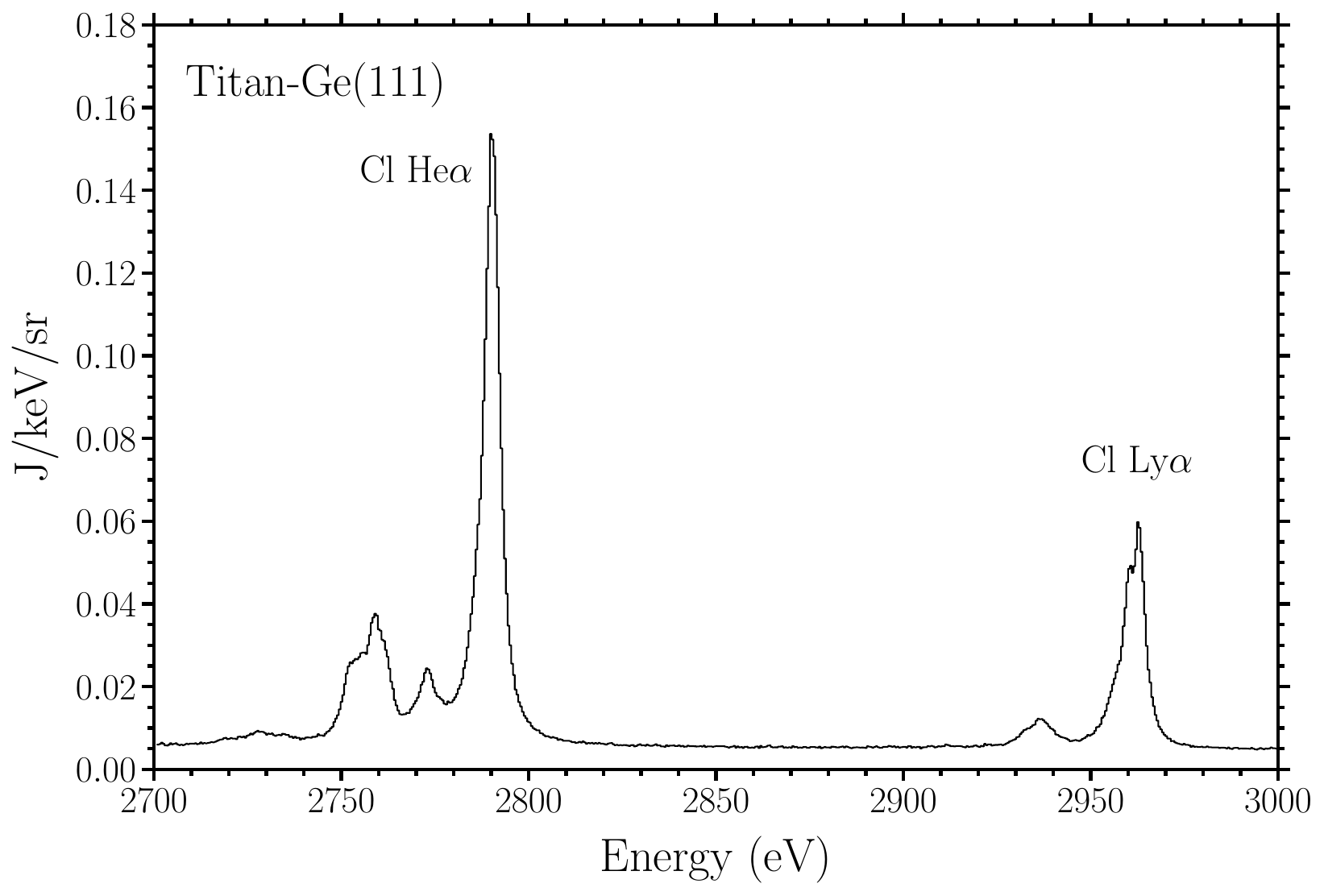}
  \end{minipage}\hfill
  \centering
  \begin{minipage}{0.48\textwidth}
    \centering
    \includegraphics[width=\linewidth]{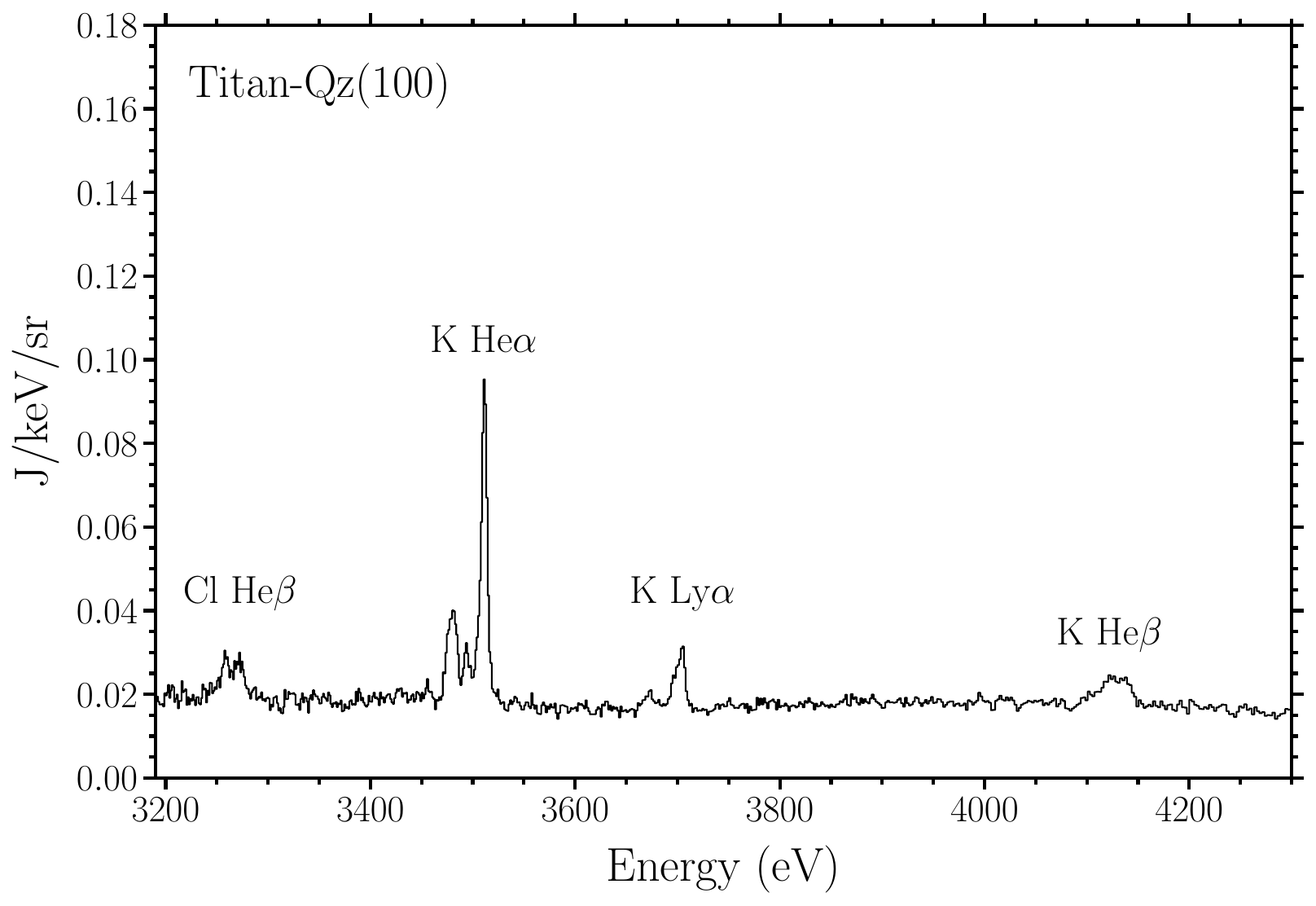}
  \end{minipage}\hfill
  \centering
  \begin{minipage}{0.48\textwidth}
    \centering
    \includegraphics[width=\linewidth]{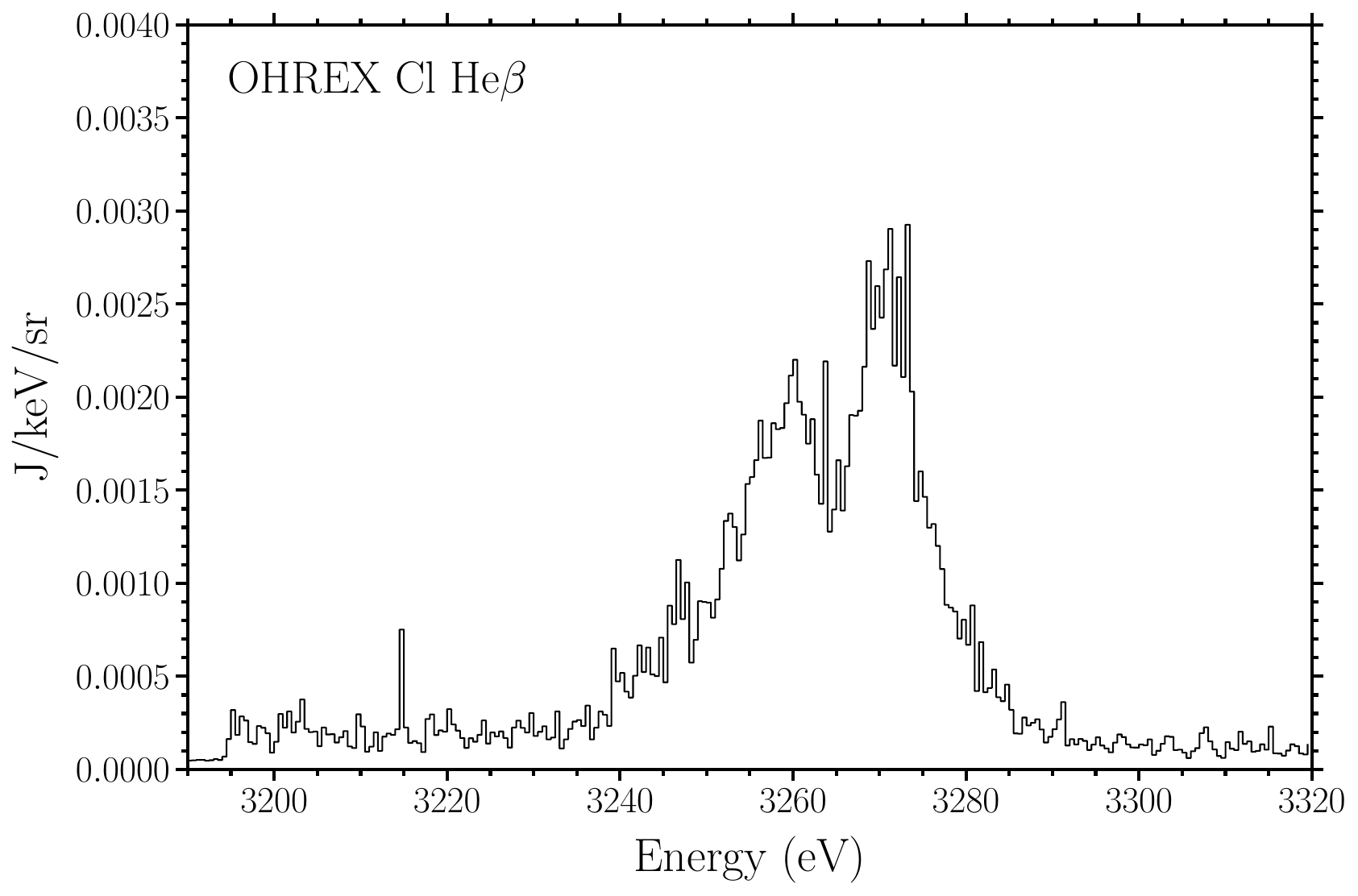}
  \end{minipage}\hfill
  \begin{minipage}{0.48\textwidth}
    \centering
    \includegraphics[width=\linewidth]{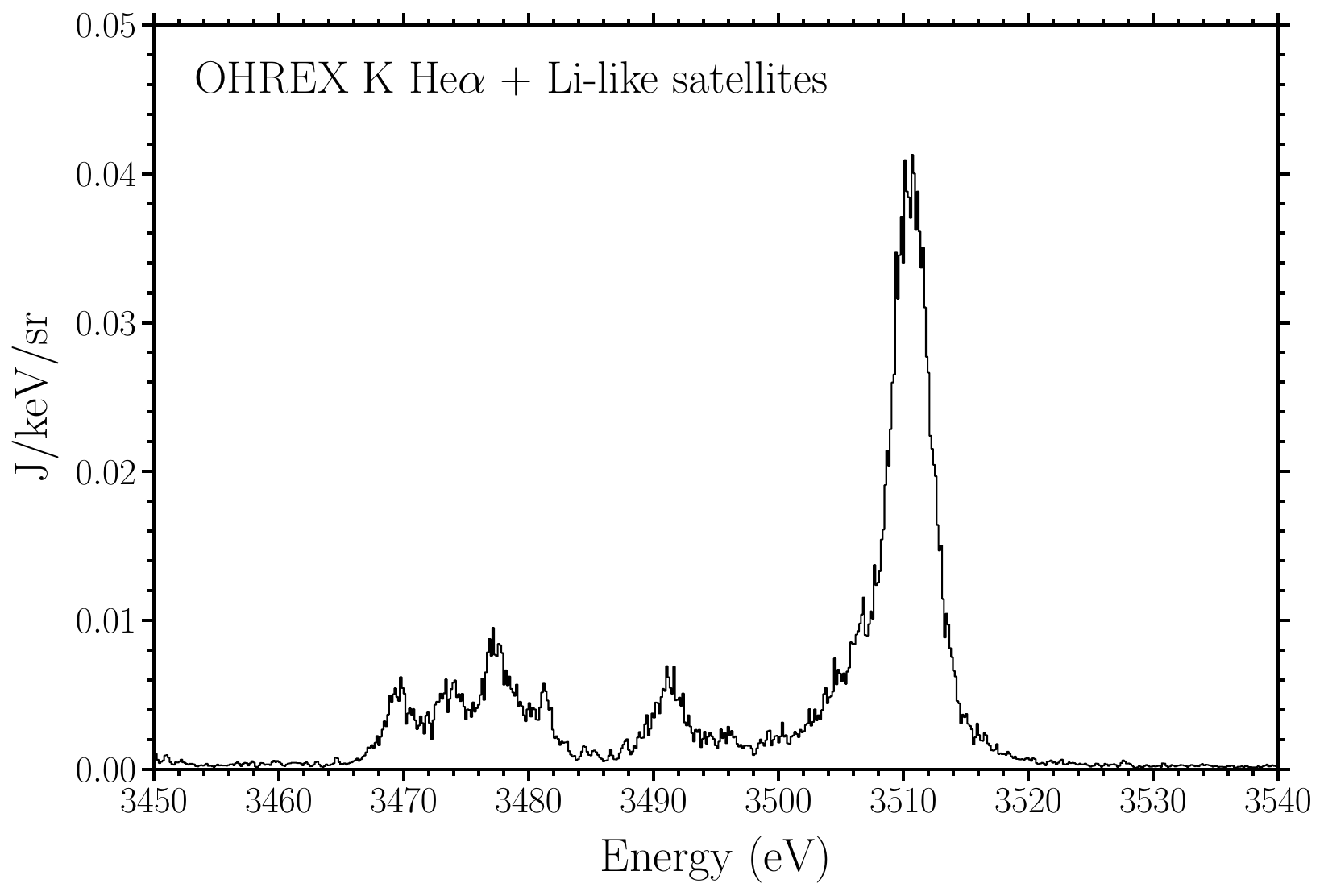}
  \end{minipage}

  \caption{Representative calibrated spectra from 2$\omega$ short-pulse experiments on a KCl buried-layer target at JLF/Titan (shot 47). Top: MKII broad-band spectrum showing H- and He-like emission from Cl and K. Middle: Titan spectrometer spectra of the Cl He$\alpha$/Ly$\alpha$ region (left) and the Cl He$\beta$/K He$\alpha$/K Ly$\alpha$ region (right). Bottom: High-resolution OHREX spectrum of the Cl$^{15+}$ He$\beta$ line (left) and the K$^{17+}$ He$\alpha$ line and its Li-like satellites (right).}
  \label{fig:shortpulse}
\end{figure*}

Figure~\ref{fig:shortpulse} presents five representative calibrated X-ray spectra from the short-pulse experiments on KCl buried-layer foil targets (shot 47). The MKII spectrometer recorded broad-band emission from H- and He-like Cl and K for time-integrated temperature and density measurements. 

The Titan spectrometer recorded Cl$^{15+}$ He$\alpha$ and Cl$^{16+}$ Ly$\alpha$ emission on a Ge(111) crystal and Cl$^{15+}$ He$\beta$, K$^{17+}$ He$\alpha$, K$^{18+}$ Ly$\alpha$, and K$^{17+}$ He$\beta$ on a Quartz(100) crystal for secondary time-integrated temperature and density measurements. The energy scale of the Ge(111) crystal was determined by fitting a linear function to pixel-versus-reference-energies using K$^{17+}$ He$\alpha$ line w and K$^{18+}$ Ly$\alpha$ lines. The energy scale of the Quartz(100) crystal was determined by fitting a third-order polynomial to pixel-versus-reference-energies using Cl$^{15+}$ He$\beta$, K$^{17+}$ He$\alpha$ line w, K$^{18+}$ Ly$\alpha$, and K$^{17+}$ He$\beta$ lines. The Quartz(100) crystal spectrum suffered from high background contamination, resulting in a much poorer signal-to-noise ratio than the Ge(111) crystal spectrum, despite sharing the same spectrometer housing. 

The OHREX spectrometer recorded Cl He$\beta$ and K He$\alpha$ complexes in high-resolution for time-integrated temperature and density measurements. Energy scales were determined by anchoring each XICSRT ray tracing dispersion to a single reference line of known rest energy. For the Cl He$\beta$ measurement, the reference point was determined by fitting a Gaussian to the higher-energy shoulder of the Cl He$\beta$ line, whose centroid we take to correspond to the unshifted Cl He$\beta$ reference energy, following Beiersdorfer \textit{et al.}~\cite{beiersdorfer:PRA:2019}. The OHREX Cl He$\beta$ was then rebinned onto a uniform 0.5~eV grid. For the K He$\alpha$ measurement, the reference point was determined by fitting a Gaussian to the K$^{17+}$ He$\alpha$ resonance line w. We note that the ray tracing dispersions are approximately linear over the short energy ranges spanned by each OHREX spectrum. 


\begin{figure}[htbp]
  \centering
  \includegraphics[width=\columnwidth]{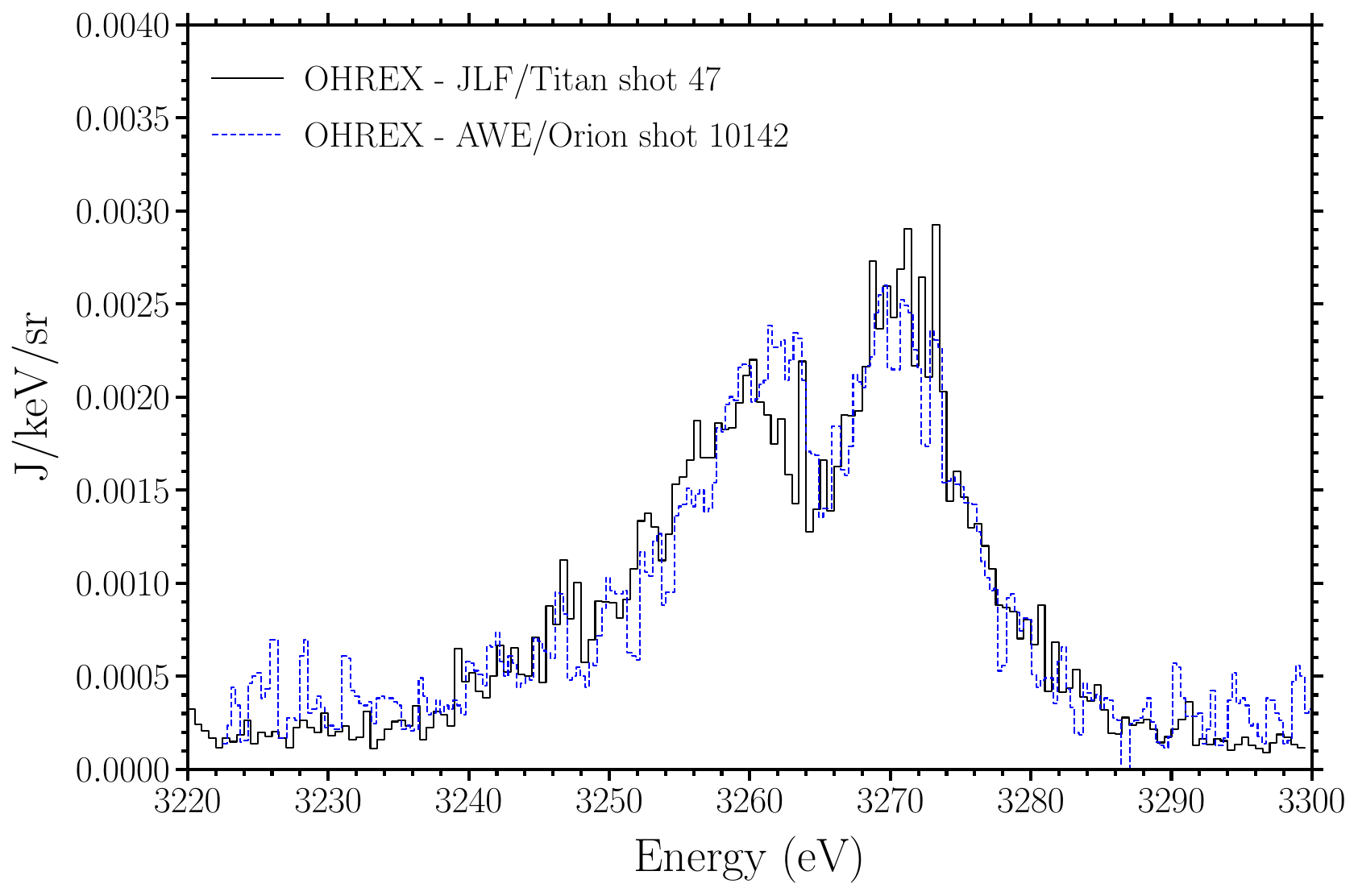}
  \caption{Comparison of high-resolution OHREX spectra of the Cl$^{15+}$ He$\beta$ line from buried-layer KCl targets, recorded at LLNL’s JLF/Titan short-pulse laser and AWE/Orion short-pulse laser. The Orion spectrum is taken from Beiersdorfer \textit{et al.}~\cite{beiersdorfer:PRA:2019} and arbitrarily scaled to match the JLF/Titan measurement.}
  \label{fig:awe}
\end{figure}

Figure~\ref{fig:awe} compares OHREX high-resolution spectra of the Cl$^{15+}$ He$\beta$ line from KCl buried-layer targets recorded at Titan and AWE's Orion Laser facility~\cite{beiersdorfer:PRA:2019}. An arbitrary scaling factor was applied to the digitized Orion data to bring the two spectra onto the same intensity scale. The line shapes are in excellent agreement, indicating similar, near-solid density plasma conditions. This demonstrates that the upgraded, frequency-doubled short-pulse Titan laser can access plasma conditions comparable to those produced by the Orion laser.

\begin{figure}[htbp]
  \centering
  \includegraphics[width=\columnwidth]{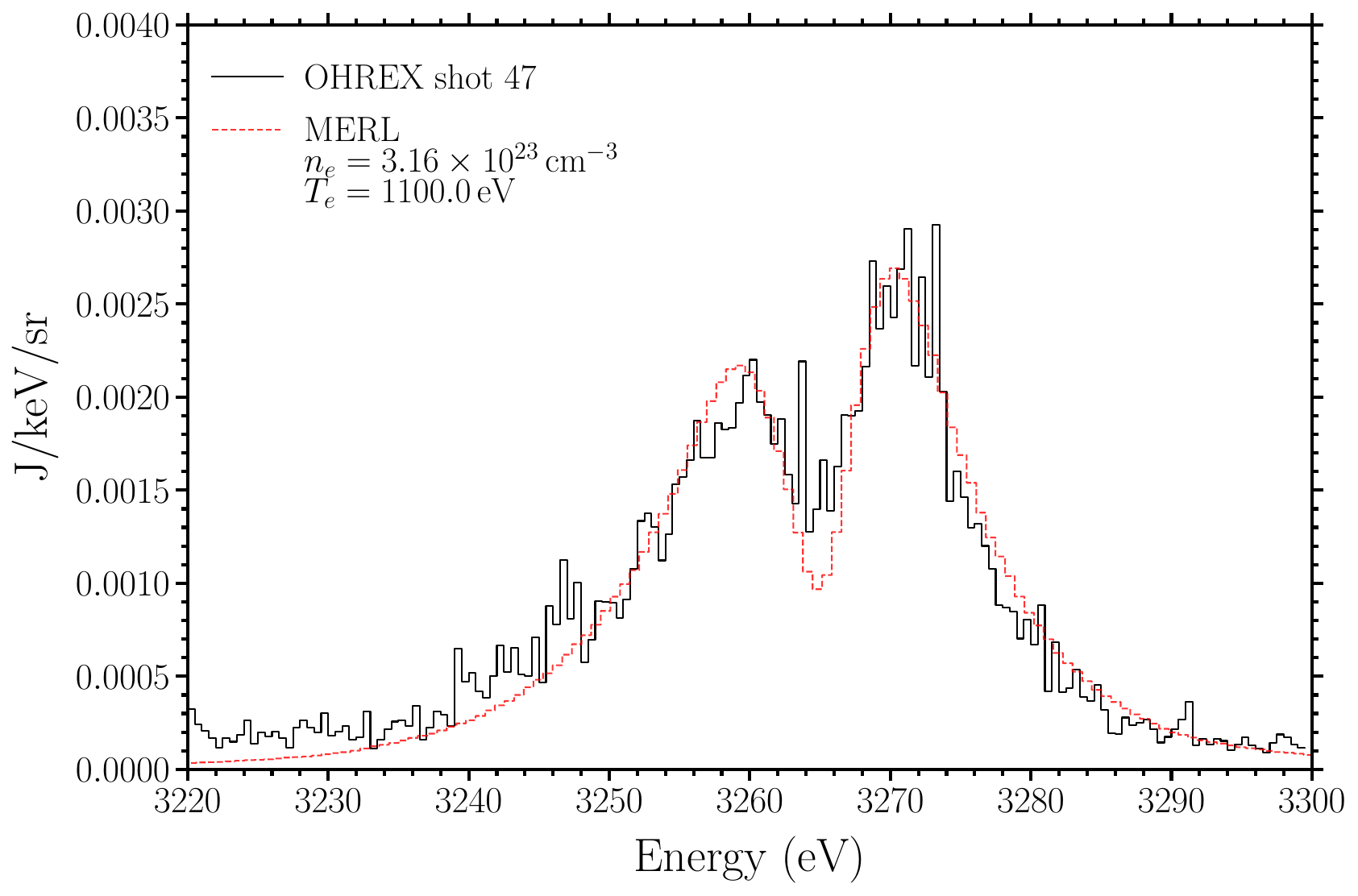}
  \caption{High-resolution OHREX Cl He$\beta$ spectrum compared with MERL line shape calculations.}
  \label{fig:merl}
\end{figure}

Figure~\ref{fig:merl} compares the high-resolution OHREX spectrum of the Cl He$\beta$ line with multi-electron radiator line shape (MERL) calculations~\cite{Woltz:PRA:1988,Mancini:CPC:1991, Mancini:HEDP:2013} at $T_e$ = 1100~eV and $n_e$ = 3.16$\times$10$^{23}$~cm$^{-3}$. A two-component least-squares fit was used to determine an energy shift and intensity scaling between the calculated and measured line shapes, yielding a shift of -3.3~eV. While the line-shape fit alone is not sufficient to independently constrain the electron temperature, it provides a strong constraint on the electron density through the Stark-broadened width of the He$\beta$ line \cite{Beiersdorfer:PP:2016, beiersdorfer:PRA:2019}.

\section{\label{sec:future}Discussion and Future Work}

Preliminary comparisons of the MKII spectrum with theoretical calculations and with short-pulse spectra obtained at Orion under similar conditions indicate a peak electron temperature as high as $\sim$ 1~keV and a near-solid electron density~\cite{Brown:OSTI:2021}. Comparing the short-pulse OHREX Cl$^{15+}$ He$\beta$ spectrum with the long-pulse spectrum (Figure~\ref{fig:longpulse}), the Li-like satellite lines visible in the long-pulse data are particularly weak in the short-pulse measurement. The weakness of these Li-like satellite lines relative to the Cl He$\beta$ line is consistent with $T_e \gtrsim$ 600~eV measured in previous experiments~\cite{Beiersdorfer:PP:2016, beiersdorfer:PRA:2019}. Additionally, the MERL fit shown in Figure~\ref{fig:merl} is also consistent with this picture. However, more work is required to accurately and quantitatively determine the temperature and density from these spectra. In particular, future analysis will need to account for temperature gradients within the X-ray-emitting plasma sampled by these time-integrated measurements, as has previously been demonstrated for time-integrated Orion spectra~\cite{MacDonald:RSI:2022}.

High-resolution spectra of dielectronic recombination (DR) Li-like satellites to He$\alpha$ and He$\beta$ complexes provide powerful plasma diagnostics and offer a route to benchmarking high-temperature, high-density atomic physics. Electron temperature can be inferred both from He$\alpha$/Ly$\alpha$ intensity ratios, which depend on the collisional-radiative ionization balance, and from DR-satellite--to--resonance-line intensity ratios, which are comparatively insensitive to the ionization balance and can therefore test the underlying collisional and dielectronic rate coefficients~\cite{Dubau:RPP:1980, Porquet:SSR:2010, Beiersdorfer:JPB:2015}. Similarly, electron density can be inferred from the Stark-broadened width of the He$\beta$ line, from the density-dependent line shift, and from the relative intensities of individual Li-like satellite lines~\cite{Beiersdorfer:PP:2016,beiersdorfer:PRA:2019}; any disagreement on the density would itself identify which model component requires refinement.

Future work will address several current experimental limitations. Direct measurements of the X-ray-emitting plasma volume using an X-ray pinhole camera will improve estimates of the emitting source size, while a newly commissioned target-chamber-center focal-spot diagnostic at JLF, equipped with a vacuum-compatible imaging system and deformable mirror, will allow the laser focal spot to be measured and optimized between shots. In addition, a more rigorous uncertainty analysis will be carried out, including independent calibrations of the image plate/scanner response and crystal reflectivities.
\section{\label{sec:conclusions}Conclusions}

We have conducted the first experiments using the recently upgraded frequency-doubled (2$\omega$) Titan short-pulse laser at LLNL. Five time-integrated spectrometers, spanning seven spectral channels, were successfully fielded across both long-pulse and short-pulse configurations on buried-layer foil targets. Calibrated spectra of H- and He-like emission from chlorine and potassium were obtained. Preliminary comparisons with theoretical calculations and with previous short-pulse measurements performed at AWE/Orion using identical spectrometers under similar laser conditions indicate peak electron temperatures as high as $\sim$ 1~keV and electron densities of $\sim$ 10$^{23}$~cm$^{-3}$. Future work will build on these initial results to produce benchmark-quality spectra for validating high-density atomic physics models and model components relevant to both astrophysical and HED communities. We are also working to make this suite of X-ray spectrometers available to the broader community, and welcome inquiries from interested collaborators.
\begin{acknowledgments}
The authors thank the staff of the Jupiter Laser Facility at Lawrence Livermore National Laboratory, especially Félicie Albert, Rick Cross, Brent Stuart, Stephen Maricle, and Matt Jamison, for their essential support, without which these experiments would not have been possible.

This work was supported by the U.S. Department of Energy Office of Science, Fusion Energy Sciences under Field Work Proposal No. SCW1836-1: LaserNetUS: Discovery Science and Inertial Fusion Energy research at the Jupiter Laser Facility, and by Lawrence Livermore National Laboratory under Contract No. DE-AC52-07NA27344.
\end{acknowledgments}

\section*{Data Availability Statement}
The data that supports the findings of this study are available from the corresponding author upon reasonable request.

\bibliography{htpd}%

\end{document}